\documentclass[aps,nobibnotes,two column,amssymb,secnumarabic,superscriptaddress,reprint]{revtex4-1}

\usepackage{amsmath}
\usepackage{amssymb}
\usepackage{bbm}
\usepackage{xcolor}
\usepackage{graphicx}
\usepackage{dcolumn}
\usepackage{bm}
\usepackage{hyperref}
\usepackage{upgreek}
\usepackage{epstopdf}

\begin{document}
\title{The effects of nucleon-nucleon short-range correlations on the inclusive electron scattering}

\author{Qinglin Niu}
\affiliation{College of Science,  China University of Petroleum (East China), Qingdao 266580, China}

\author{Jian Liu}\email{liujian@upc.edu.cn}
\affiliation{College of Science,  China University of Petroleum (East China), Qingdao 266580, China}
\affiliation{The Key Laboratory of High Precision Nuclear Spectroscopy, Insititute of Modern Physics, Chinese Academy of Sciences}
\affiliation{Guangxi Key Laboratory of Nuclear Physics and Nuclear Technology, Guangxi Normal University}

\author{Yuanlong Guo}
\affiliation{College of Science,  China University of Petroleum (East China), Qingdao 266580, China}

\author{Chang Xu}
\affiliation{Department of Physics, Nanjing University, Nanjing 210093, China }

\author{Mengjiao Lyu}
\affiliation{College of Science, Nanjing University of Aeronautics and Astronautics, Nanjing 210016, China}

\author{Zhongzhou Ren}
\affiliation{School of Physics Science and Engineering, Tongji University, Shanghai 200092, China}

\begin{abstract}
The nucleon-nucleon short-range correlation (\textit{NN}-SRC) is one of the key issues of nuclear physics, which typically manifest themselves in high-momentum components of the nuclear momentum distributions. In this letter, the nuclear spectral functions based on the axially deformed relativistic mean-field model are developed to involve the \textit{NN}-SRC. With the spectral functions, the inclusive electron scattering $ (e,e^{\prime}) $ cross sections are calculated within the PWIA framework, including the quasi-elastic (QE) part and $ \Delta $ production part. Especially in the $ \Delta $ production region, we reconsider the electromagnetic structures of the nucleon resonance $ \Delta $(1232) and the scattering mechanisms, thereby the theoretical calculations are improved effectively and the cross sections are well consistent with the experimental data. The theoretical $ (e,e^{\prime}) $ cross sections are further divided into \textit{NN}-SRC and mean-field contributions. It is found that, at the kinematics $ 0.5 \,{\rm GeV}^{2}<Q^{2}<1 \,{\rm GeV}^{2} $, the QE peak and $ \Delta $ production peak not only reflect the mean-field structure, but also are sensitive to the \textit{NN}-SRC information. Finally,  we provide a new method to extract the strengths of \textit{NN}-SRC from experimental cross sections for selected nuclei at the suitable kinematics.

\end{abstract}
\maketitle
\emph{Introduction. }
Many-body Fermi systems including the short-range correlation (SRC) are common in nature, such as nuclear matter system and finite nuclei. The information of nucleon-nucleon short-range correlation (\textit{NN}-SRC) is useful to answer several key issues about the properties of matter at high densities, such as neutron stars and relativistic heavy-ion collisions\cite{PhysRevC.91.025803,li2018nucleon}. The research of the EMC effect would benefit from the details of \textit{NN}-SRC, which  describes  the modification of the quark–gluon structure of a nucleon bound in an atomic nucleus by the surrounding nucleons \cite{hen2017nucleon}.

The information of \textit{NN}-SRC can be obtained by new experimental data on electron scattering off nuclei. Corresponding experiments have been carried out at JLab\cite{duer2019direct,Duer2018probing,Schmookler2019Modified}, which are categorized into two groups: the exclusive electron scattering $ (e, e^{\prime}p) $ and the inclusive electron scattering $ (e,e^{\prime}) $. Previous measurements of $ (e, e^{\prime}p) $ experiments on $ ^{12} $C to $ ^{208} $Pb\cite{subedi2008probing,hen2014momentum} pointed out that the \textit{np}-SRC pairs are approximately 20 times as many as \textit{pp}-SRC and \textit{nn}-SRC pairs, which originate from the tensor part of the nuclear force \cite{schiavilla2007tensor}. The $ (e,e^{\prime}) $ experiments have shown that SRC pair nucleons are shifted from low-momentum states to high-momentum states \cite{degli2015medium}. According to the $ (e,e^{\prime}) $ experiments at SLAC and JLab\cite{frankfurt1993evidence,egiyan2006measurement,fomin2012new}, the ratios of $ (e,e^{\prime}) $ cross sections on heavy nuclei to those of the deuteron exhibit a plateau, at the squared four-momentum transfer $ Q^{2}>1.4 \rm \,GeV^{2} $ and large Bjorken scaling variable $ 1.5 < x_{B}  <2 $. The plateau indicates that the high-momentum components of different nuclei have a similar shape, and the value of the plateau provides a chance to extract the proportion of high-momentum nucleons.  As the development of experiment mentioned above, new theoretical works in the field of electron scattering are also stimulated to explain the experimental phenomenon and extract the information of nuclear structure information.

The plane-wave impulse approximation (PWIA) is a convenient theory widely applied to explain the quasi-elastic (QE) scattering and $ {\Delta} $ production scattering\cite{benhar2008inclusive}. Within the PWIA framework, the spectral function can be introduced into the descriptions of the $ (e,e^{\prime}) $ scattering reaction. The spectral function $ S(\textbf{p}, E) $ is the probability of finding a nucleon with given momentum $ \textbf{p} $ and removal energy $ E $ in nuclei\cite{benhar2008inclusive}. The $ (e,e^{\prime}) $ cross sections can be obtained by the two physical quantities: the spectral functions $ S(\textbf{p}, E) $ representing the nuclear structure, and the elementary cross section $ \sigma_{e N} $ describing the process of an electron scattered by an off-shell nucleon. It is a key problem for $ (e,e^{\prime}) $ scattering to calculate the elementary cross section, which depends on the information the electromagnetic structure of the nucleon. For the QE region, the cross sections have been well reproduced at the quantitative level. However, because of the uncertainties associated with the form factors in the $ {\Delta} $ production region, there are disagreements between theory and experimental data, and the $ {\Delta} $ production region still needs further studies\cite{benhar2006estimates}.

 Another key problem in $ (e,e^{\prime}) $ scattering is accurate calculations of spectral function. The theoretical calculations of the spectral function $ S(\textbf{p}, E) $ can be assumed to consist of the MF part $ S_{\rm MF}(\textbf{p}, E) $ and the \textit{NN}-SRC part $ S_{\rm corr}(\textbf{p}, E) $\cite{benhar2008inclusive,kulagin2006global,degli2015medium}. The MF part $ S_{\rm MF}(\textbf{p}, E) $ gives the descriptions that the nucleons move in well-defined quantum states, under the effect of an average field created by their interactions. Compared with the $ S_{\rm MF}(\textbf{p}, E) $, the \textit{NN}-SRC part $ S_{\rm corr}(\textbf{p}, E) $ provides supplementary description of nucleons with high momentum and high removal energy above the Fermi surface, which are caused by tensor attraction and short-range repulsion. The \textit{ab initio} method\cite{pieper2001quantum,wan2020finite} is a fundamental method to describe the \textit{NN}-SRC of light nuclei, but it is still difficult to be applied to the medium and heavy nuclei. Because the \textit{NN}-SRC is unaffected by surface and shell effects\cite{hen2017nucleon}, the $ S_{\rm corr}(\textbf{p}, E) $ of the finite nuclei can be evaluated from the rescaling the strength of \textit{NN}-SRC of deuteron\cite{antonov2002nucleon}.

 In this letter, the $ (e,e^{\prime}) $ scattering theory is linked with the microscopic nuclear structure model, and the influences of \textit{NN}-SRC on the $ (e,e^{\prime}) $ cross sections are studied. In the calculations of $ {\Delta} $ production scattering, the appropriate form factors are used, in addition, the conservation of energy and momentum in the scattering process and the decay width of $ \Delta $(1232) are also considered. Therefore, the calculation results have been significantly improved. On the basis of theoretical results, we study the effect of the \textit{NN}-SRC part and the mean-field part on the $ (e,e^{\prime}) $ cross sections, respectively. Furthermore, this paper analyzes the sensitivity of the inclusive electron scattering to the \textit{NN}-SRC at the different kinematics. Finally, a new method is proposed to extract the strength of \textit{NN}-SRC from the experimental cross sections.

~\

\emph{Formulas. }\label{s3}
 The calculations of spectral function in this paper are divided into two categories: the MF part $ S_{\rm MF}(\textbf{p}, E) $ and the $ S_{\rm corr}(\textbf{p}, E) $
\begin{equation}
	S(\mathbf{p}, E)=S_{\mathrm{MF}}(\mathbf{p}, E)+S_{\text {corr }}(\mathbf{p}, E),
	\label{eq:one}
\end{equation}
with the normalization requirement  $ \int d^{3} p dES(\mathbf{p}, E)=A $. In this paper, the MF parts are calculated by the axially deformed relativistic mean-field (RMF) model. At low energy $ E $ and low momentum $ \textbf{p} $, the MF part $ S_{\rm MF}(\textbf{p}, E) $ is dominated by the  single-particle properties,
\begin{equation}
	S_{\mathrm{MF}}(\mathbf{p}, E)=\sum_{i} C_{i}\left(\left|f_{i}(\mathbf{p})\right|^{2}+\left|g_{i}(\mathbf{p})\right|^{2}\right) L_{i}\left(E-E_{i}\right),
	\label{eq:two}
\end{equation}
where $ f_{i}(\mathbf{p}) $ and $ g_{i}(\mathbf{p}) $ are the two-dimensional Dirac spinors in momentum space for the single-particle state $ i $, and $ C_{i} $ is the corresponding occupation number of the single particle state \textit{i}.  The finite width in energy dependence can be described by Lorentzian function $ L_{i} $\cite{ivanov2019realistic}.

At the region of high energy $ E $ and high momentum $ \textbf{p} $, the \textit{NN}-SRC part $ S_{\rm corr}(\textbf{p}, E) $ plays a leading role because the strong dynamical \textit{NN}-SRC leads to virtual scattering processes exciting the nucleons to the states above the fermi surface. The $ S_{\rm corr}(\textbf{p}, E) $ is determined by ground state configurations with the \textit{NN}-SRC pair and the $ (A-2) $ nucleon residual system. The \textit{NN}-SRC pair has a high relative momentum $ \textbf{p}_{\mathrm{rel}} $ and a low center-of-mass momentum $ \textbf{p}_{\mathrm{CM}} $. At the region of high $ \textbf{p} $, the \textit{NN}-SRC momentum distribution originates from the relative momentum distribution $ n_{\text {corr }}(\textbf{p})=n_{\text {rel }}(\textbf{p}) $, because of the low center-of-mass momentum. Assuming the  $ n_{\mathrm{CM}}\left(\textbf{p}_{\mathrm{CM}}\right)=(\alpha / \pi)^{3 / 2} \exp \left(-\alpha \textbf{p}_{\mathrm{CM}}^{2}\right) $, the \textit{S}$ _{\rm corr} $(\textbf{p},\textit{E}) can be obtained by the integral of $  \textbf{p}_{\mathrm{CM}} $\cite{kulagin2006global}
\begin{equation}
	\begin{aligned}
		S_{\text {corr }}(\mathbf{p}, E)=& n_{\text {corr }}(\mathbf{p}) \frac{m}{|\mathbf{p}|} \sqrt{\frac{\alpha}{\uppi}} \\
		& \times\left[\exp \left(-\alpha \mathbf{p}_{\min }^{2}\right)-\exp \left(-\alpha \mathbf{p}_{\max }^{2}\right)\right],
	\end{aligned}
	\label{eq:three}
\end{equation}
where the $ {\alpha} = 3 /\left(4\left\langle\mathbf{p}^{2}\right\rangle(A-2) /(A-1)\right) $. The $ \textbf{p}_{\mathrm{min}} $ and the $ \textbf{p}_{\mathrm{max}} $ are the lower and upper limits of the $ \textbf{p}_{\mathrm{CM}} $.

There are many microscopic or phenomenological methods to calculate the $ n_{\rm corr} (\textbf{p}) $, and the light-front dynamics (LFD) method is adopted in this paper. In the LFD method, the $ n_{\rm corr} (\textbf{p}) $ can be obtained by rescaling the \textit{NN}-SRC part of the precise momentum distribution of deuteron. The momentum distribution can be expressed as follows:
\begin{equation}
	n_{\text {corr }}^{\tau}(\mathbf{p})=N_{\tau} \tau C_{A}\left[n_{2}(\mathbf{p})+n_{5}(\mathbf{p})\right],
	\label{eq:four}
\end{equation}
where the two components $ n_{2} (\textbf{p}) $ and $ n_{5} (\textbf{p}) $ reflect of deuteron, which are deduced from LFD wave functions\cite{carbonell1995relativistic}. The $ n_{2} (\textbf{p}) $ term is mainly induced by the tensor force, the  $ n_{5} (\textbf{p}) $ term is mainly induced by the $ {\pi} $-exchange\cite{wang2021nucleon}. And the scaling factor $ C_{A} $ is the ratio of high-momentum components between other nuclei and deuteron. For different nuclei, the scaling factor $ C_{A} $ represents the strength of \textit{NN}-SRC, which needs to be extracted by electron scattering.

During the inclusive electron scattering $ (e,e^{\prime}) $ process, the incident electrons are interacted with the target nucleus and scattered off from four-momentum  $ k \equiv\left(E_{\mathbf{k}}, \mathbf{k}\right) $ to $ k^{\prime} \equiv\left(E_{\mathbf{k}^{\prime}}, \mathbf{k}^{\prime}\right) $, The four-momentum transfer of scattered electrons is $ q \equiv(\omega, \mathbf{q}) $. Neglecting the final-state interactions, the double differential cross sections can be written as\cite{benhar2008inclusive}
\begin{equation}
	\frac{d^{2} \sigma}{d \Omega d E_{\mathbf{k}^{\prime}}}=\frac{\alpha^{2}}{Q^{4}} \frac{E_{\mathbf{k}^{\prime}}}{E_{\mathbf{k}}} L_{\mu \nu} W^{\mu \nu},
	\label{eq:five}
\end{equation}
where $ {\alpha} $ is the fine structure constant and the squared four-momentum transfer $ Q^{2}=-q^{2}=\omega^{2}-\mathbf{q}^{2} $. In Eq.~(\ref{eq:five}), the leptonic tensor $ L_{\mu \nu} $ is completely determined by the electron kinematics, and the nuclear tensor $ W^{\mu \nu} $ includes all the information of target nuclear
\begin{equation}
	W^{\mu \nu}=\sum_{X}\left\langle 0\left|J^{\mu}\right| X\right\rangle\left\langle X\left|J^{\nu}\right| 0\right\rangle \delta^{(4)}\left(p_{0}+q-p_{X}\right),
	\label{eq:six}
\end{equation}
where $ J^{\mu} $ is nuclear currents, $ p_{0} $ and $ p_{X} $ are the four-momentums of hadronic initial-states and final-states. In IA scheme, the scattering process is considered the incoherent sum of elementary scattering reaction involving only one nucleon. Therefore, the nuclear tensor can be described as
\begin{equation}
	W^{\mu \nu}=\sum_{i} \int d^{3} p d E w_{i}^{\mu \nu}(\tilde{q})\left(\frac{m}{E_{\mathbf{p}}}\right) S(\mathbf{p}, E).
	\label{eq:seven}
\end{equation}
The nucleon tensor $ w_{i}^{\mu \nu} $ reflects the electromagnetic interactions of a bound nucleon carrying momentum \textbf{p}.

Combining with  Eqs.~(\ref{eq:five}) and (\ref{eq:seven}), the $ (e,e^{\prime}) $ cross sections can be written in the form
\begin{equation}
	\begin{aligned}
		\frac{d^{2} \sigma}{d \Omega d E_{\mathbf k^{\prime}}}=& \int d^{3} p d E \left[S_{p}(\mathbf{p}, E) \frac{d^{2} \sigma_{e p}}{d \Omega d E_{\mathbf k^{\prime}}}\right.\\
		&\left.+S_{n}(\mathbf{p}, E) \frac{d^{2} \sigma_{e n}}{d \Omega d E_{\mathbf k^{\prime}}}\right] \delta\left(\omega-E+m-E_{\left|\mathbf{p}+\mathbf{q}\right|}\right),
	\end{aligned}
	\label{eq:eight}
\end{equation}
the elementary cross section $ {d^{2} \sigma_{e N}}/{d \Omega d E_{\mathbf k^{\prime}}} $ represents the scattering of an electron by a nucleon for the QE and the $ \Delta $ production process\cite{benhar2008inclusive}. We calculate the $ {d^{2} \sigma_{e N}}/{d \Omega d E_{\mathbf k^{\prime}}} $  using the form factors in QE process\cite{benhar2008inclusive} and the $ \Delta $ production process\cite{amaro1999relativistic}, which reflect the electromagnetic structure of the nucleon and the $ \Delta $(1232). For the integration process, we determine the limits of integration by considering the conservation of energy and momentum in the scattering process.

In particular, for the $ \Delta $ production process, the Lorentzian shape $ \Delta $ width is used as a substitute for the energy-conserving delta function in Eq.~(\ref{eq:eight}) to produce a broadening of the $ \Delta $ peak and correspondingly a decrease of the strength
\begin{equation}
	\delta \rightarrow \frac{1}{\pi} \frac{{\mit\Gamma} \left(W\right) / 2}{\left(W-m_{\Delta}\right)^{2}+{\mit\Gamma} \left(W\right)^{2} / 4},
	\label{eq:nine}
\end{equation}
where the parameter $ W $ is the invariant mass. The parameter $  \mit\Gamma (W) $ is the decay width of $ \Delta $(1232)\cite{amaro1999relativistic}. The integration interval of \textit{W} goes from threshold to the maximum value allowed in the Fermi gas model $ m+m_{\pi}<W<\sqrt{\left(E_{F}+\omega\right)^{2}-\left(q-p_{F}\right)^{2}} $.

~\

\emph{Momentum distribution and spectral function. }
 The nucleon momentum distributions $ n(p) $ and the nuclear spectral functions $ S(p, E) $ are calculated with the formulas above. The corresponding nuclear wave functions in momentum space are obtained from the axially deformed RMF model with the NL3* parameter set. The correlation part $ S_{\rm corr}(p, E) $ is obtained from the LFD method with the strength $ C_{A}=4.5$  in Eq.~(\ref{eq:four}).

\begin{figure}
	\includegraphics[width=0.43\textwidth]{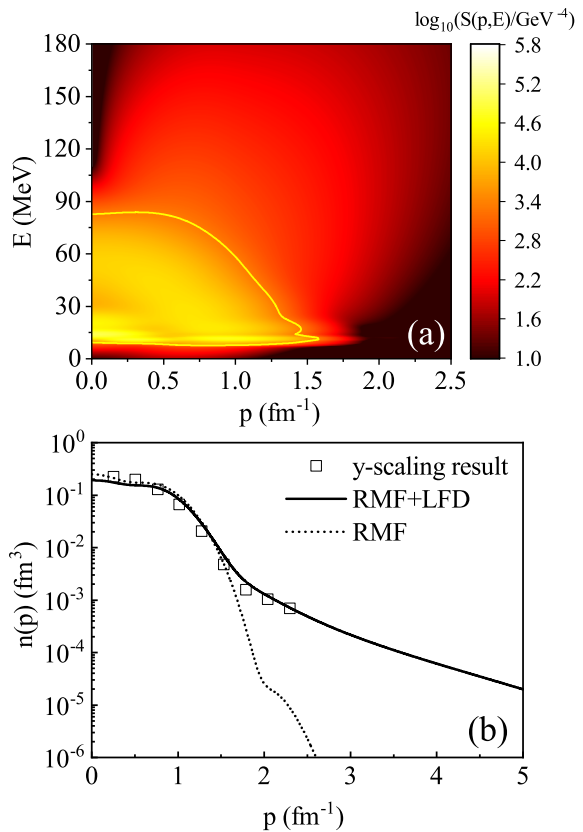}
	\caption{\label{fig1} (Color online) Momentum distribution $ n(p) $ and spectral functions $ S(p, E) $ of $ ^{56} $Fe for the deformation $ \beta=0.17 $ calculated from the deformed RMF model with the LFD method. In Fig.~\ref{fig1}(a), the logarithm of $ S(p, E) $ is presented to highlight the \textit{NN}-SRC part, and the region enclosed by a curve describes the mean-field part. The open squares represent the $ n(p) $ obtained from y-scaling analyses on $ (e,e^{\prime}) $ cross sections\cite{Ciofi1991y-scaling}.  }
\end{figure}

In Fig.~\ref{fig1}(a), we present the logarithm of spectral function $ S(p, E) $ of $ ^{56} $Fe to highlight the \textit{NN}-SRC part. We observed that in the region of high removal energy $ (E>100{\rm \, MeV}) $ and high momentum $ (p > 1.5 {\rm \,fm }^{-1} ) $, the spectral function is mainly from the contributions of \textit{NN}-SRC, which has no shell structure and shows in a smooth ridge. Different from the \textit{NN}-SRC part, one can clearly distinguish the different orbits from the mean-field $ S_{\rm MF}(p, E) $ in the region of $ E <90 {\rm \,MeV} $ and $ p < 1.5 {\rm\, fm }^{-1}  $ (region enclosed by a curve). The corresponding nucleon momentum distribution $ n(p) $ of $ ^{56} $Fe are calculated by $ S(p, E) $ and presented in Fig.~\ref{fig1}(b). The momentum distribution extracted from the $ (e,e^{\prime}) $ cross sections and analyzed in terms of y-scaling in Ref.~\cite{Ciofi1987Nucleon,Ciofi1991y-scaling} are also provided in this figure for comparison. From Fig.~\ref{fig1}(b), the mean-field calculations only provide good descriptions on $ n(p) $ below the Fermi momentum $ p_{ F}  \approx 1.39 {\rm \,fm} ^{-1} $. For $ p>p_{F} $, the $ n(p) $ from the pure mean-field calculations decrease rapidly and deviate from the $ n(p) $ from the y-scaling analyses. By considering the \textit{NN}-SRC contributions, the value of $ n(p) $ on the tail is enhanced and consistent with the result of the y-scaling analyses.

~\

\begin{figure}
	\includegraphics[width=0.43\textwidth]{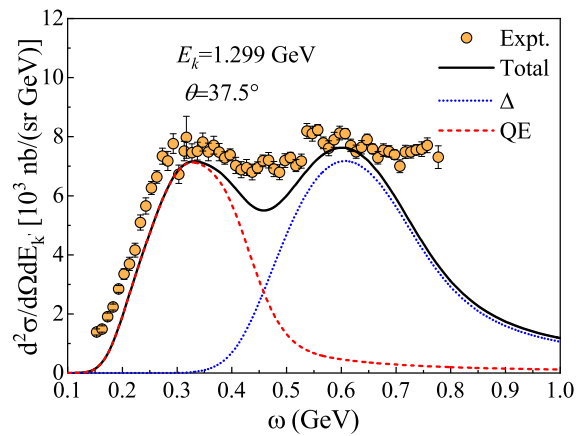}
	\caption{\label{fig2} (Color online) The total $ (e,e^{\prime}) $ cross sections of $ ^{56} $Fe calculated by PWIA method at the kinematics $ E_{k}= 1.299{\rm\, GeV} $ and $ \theta= 37.5^{\circ} $, where the spectral function is from RMF+LFD calculations. The experimental data are from Ref.~\cite{sealock1989electroexcitation}. Dotted lines: the cross sections of the $ \Delta $ production peaks; dashed lines: the cross sections of the QE peaks. }
\end{figure}

\emph{Inclusive electron scattering cross sections. }
In Fig.~\ref{fig2} we present theoretical $ (e,e^{\prime}) $ cross sections of $ ^{56} $Fe from the spectral function of Fig.~\ref{fig1}(a). The theoretical results of the inclusive electron scattering can be divided into the QE peak and $ \Delta $ production peak. Good agreements with the experimental data are obtained in the total theoretical cross sections. Especially for the $ \Delta $ production region, we observed that the location and the value of the experimental peak can be reproduced well by the Eqs.~(\ref{eq:eight}) and~(\ref{eq:nine}). The calculations of $ \Delta $ production cross sections rely on the nucleon structure functions including the magnetic, electric and Coulomb excitation form factors. Thus, the $ \Delta $ production peak contains information about the nucleon's excited states and the electromagnetic structure of $ \Delta $(1232) baryon, which is beneficial for understanding strong interactions in the domain of quark confinement.

Both the QE scattering and $ \Delta $ production scattering follow the energy and momentum conservations. The QE peak locates at the energy transfer $ \omega_{QE} = Q^{2} /2m =0.33{\rm\, GeV} $, which reflects the process of the electron scattered by a free nucleon\cite{wang2021global}. For the $ \Delta $ production scattering, the peak locates at the $ \omega_{\Delta}=m_\mathrm{\Delta}-m+\omega_\mathrm{QE}=0.62\, \mathrm{GeV} $, which corresponds to the excitation of a free nucleon to form a $ \Delta $(l232).

~\

\begin{figure}
	\includegraphics[width=0.43\textwidth]{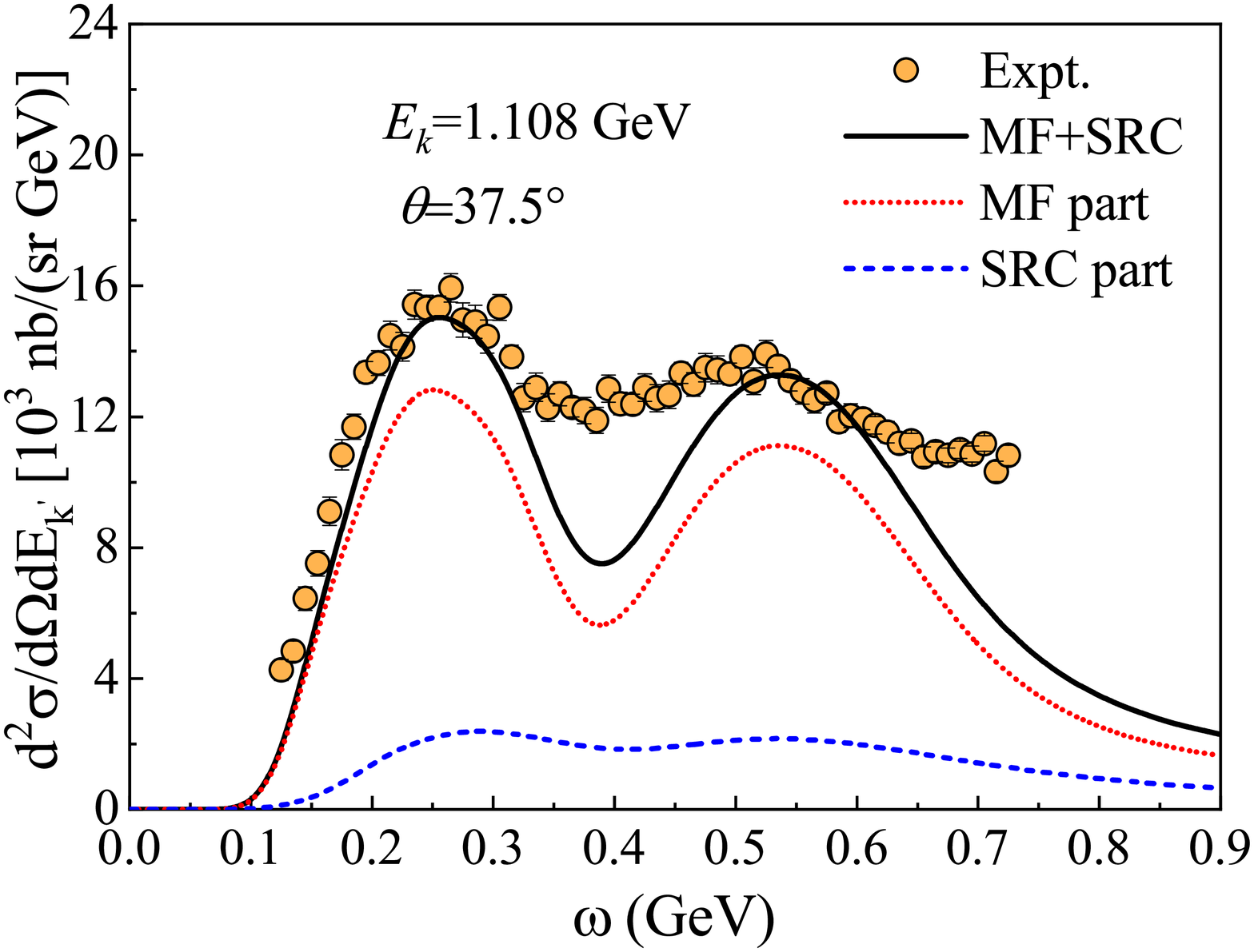}
	\caption{\label{fig3} (Color online) The $ (e,e^{\prime}) $ cross sections of $ ^{56} $Fe at the kinematics $ E_{k}= 1.108{\rm\, GeV} $ and $ \theta= 37.5^{\circ} $, calculated with the $ S(p, E) $ of Fig.~\ref{fig1}(a). Solid line: the total inclusive cross sections. Dotted line: the contributions of MF part. Dashed line: the contributions of \textit{NN}-SRC par. The experimental data are from the Ref.~\cite{chen1991longitudinal}.}
\end{figure}

\emph{Effects of NN-SRC. }
In this part, we further separate the $ (e,e^{\prime}) $ cross sections into the MF part and the \textit{NN}-SRC part. The results of $ ^{56} $Fe are exhibited in Fig.~\ref{fig3}, calculated with the kinematics $ E_{k} =1.108{\rm\, GeV}$ and $ \theta = 37.5^{\circ}$. For the strength of \textit{NN}-SRC, the corresponding proportion of high-momentum nucleons is $Y=21.4\%$ in total nucleon number. However, in Fig.~\ref{fig3} the contributions of \textit{NN}-SRC part to the total cross sections are about $16\%$. It means that parts of the \textit{NN}-SRC nucleons with high $ E $ and $ p $ cannot take part in the $ (e,e^{\prime}) $ reaction,  because of the conservation of momentum and energy. With the increase of incident energy and angle, the contributions of \textit{NN}-SRC nucleons to the total cross sections are enhanced. Therefore, it is suitable to extract the strength of \textit{NN}-SRC from the $ (e,e^{\prime}) $ reaction at the kinematics with high $ Q^{2} $.

It was pointed out that the ratio of the $ (e,e^{\prime}) $ cross sections  per nucleon of nucleus A to that of deuterium  have a plateau\cite{frankfurt1993evidence,egiyan2006measurement,fomin2012new}, in the range of the Bjorken scaling variable $ 1.5<x_{B}<2 $ and the squared four-momentum transfer $ Q^{2}>2{\rm\, GeV^{2}} $. At this condition, the $ (e,e^{\prime}) $ cross sections are mainly from the \textit{NN}-SRC nucleons, and the contributions of mean-field nucleons can be negligible because of the energy and momentum conservations. From the analyses of Fig.~\ref{fig3}, we also found that the kinematics with high $ Q^{2} $ are suitable to extract the strength of \textit{NN}-SRC, which agrees with previous studies.
\begin{figure*}
	\includegraphics[width=0.9\textwidth]{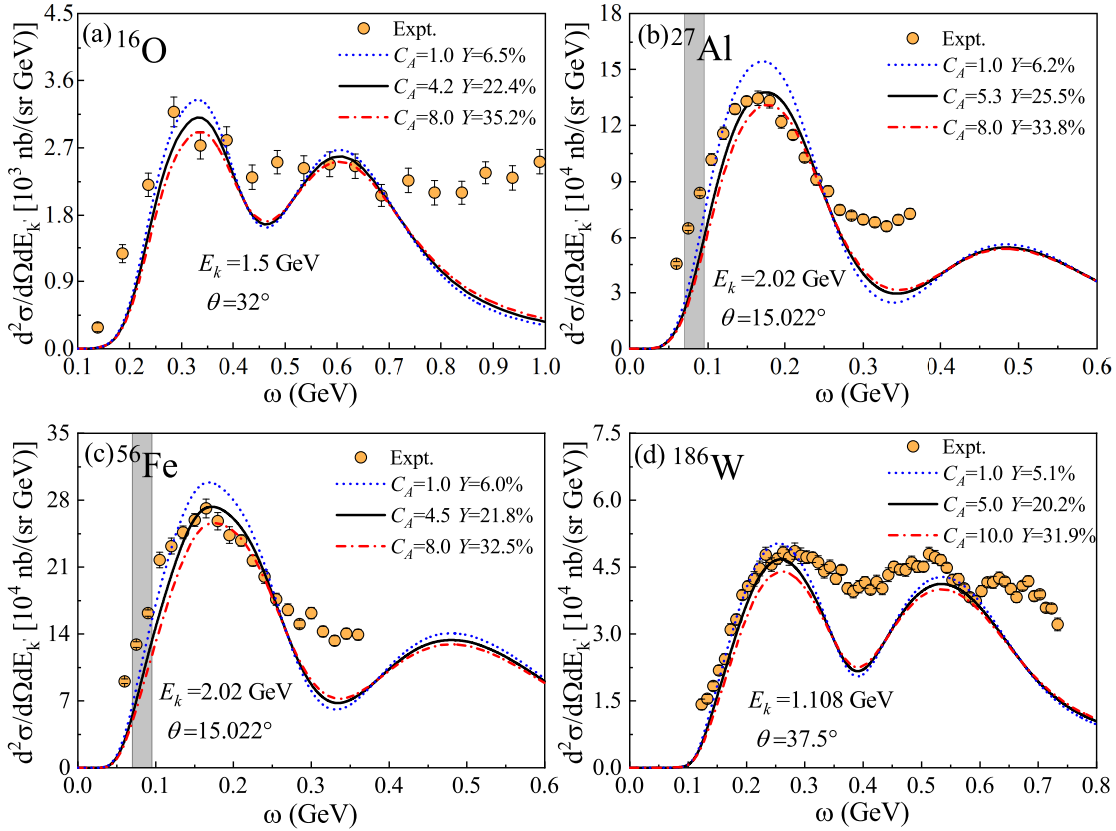}
	\caption{\label{fig4}(Color online) (a) The inclusive cross sections of $ ^{16} $O at the kinematic $ E_{k}= 1.5{\rm\, GeV} $ and $ \theta= 32^{\circ} $ for different \textit{NN}-SRC strengths, calculated from the spectral function from RMF+LFD method. (b) Same as (a) but for $ ^{27} $Al at the kinematic  $ E_{k}= 2.02{\rm\, GeV} $ and $ \theta= 15.022^{\circ} $. (c) Same as (a) but for $ ^{56} $Fe at the kinematic $ E_{k}= 2.02{\rm\, GeV} $ and $ \theta= 15.022^{\circ} $. (d) Same as (a) but for $ ^{186} $W at the kinematic $ E_{k}= 1.108{\rm\, GeV} $ and $ \theta= 37.5^{\circ} $. The shaded areas in (b) and (c) represent the region of the Bjorken scaling variable $ 1.5<x_{B}<2 $. The experimental data are from Refs.~\cite{sealock1989electroexcitation,anghinolfi1996quasi,day1993inclusive}.}
\end{figure*}

~\

\emph{Extract the strength of the NN-SRC. }
In Fig.~\ref{fig4}, we present the inclusive cross sections of $ ^{16} $O, $ ^{27} $Al, $ ^{56} $Fe, and $ ^{186} $W calculated by the spectral function for the different strength of \textit{NN}-SRC. To extract the information of \textit{NN}-SRC form nuclei, we choose the squared four-momentum transfer ranges from $ 0.5 $ to $ 1 {\rm\, GeV^{2}} $, and the corresponding kinematics are given in Fig.~\ref{fig4}. From Fig.~\ref{fig4}, one can observe there are noticeable changes of $ (e,e^{\prime}) $ cross sections near the QE and $ \Delta $ production peaks for different correlation strength. For $ ^{16} $O in Fig.~\ref{fig4}(a), the correlation strengths are $ C_{A} =1.0$, $ 4.2$, and $ 8.0 $, and the high-momentum nucleons account for $6.5\%$, $22.4\%$, and $35.2\%$ in total nucleon number, respectively. Comparing the theoretical $ (e,e^{\prime}) $ cross sections with the experimental data, the strengths of \textit{NN}-SRC are constrained to be $ C_{A}=4.2 $ for $ ^{16} $O, which indicates the correlated nucleons contribute to about $22.4\%$ of the total nucleon number. For $ ^{27} $Al, $ ^{56} $Fe, and $ ^{186} $W, the corresponding analysis are also carried out. Finally, we obtain that the strengths of \textit{NN}-SRC are $ C_{A}=5.3 $ for $ ^{27} $Al, $ C_{A}=4.5 $ for $ ^{56} $Fe, and $ C_{A}=5.0 $ for $ ^{186} $W. The percentages of high-momentum nucleons in total nucleon number are $Y=25.5\%$ for $ ^{27} $Al, $Y=21.4\%$ for $ ^{56} $Fe, and $Y=20.2\%$ for $ ^{186} $W. In this analysis, we found that the correlated terms contribute to about $20\%$ of the total nucleon in medium and heavy nuclei, which verifies the \textit{ab initio} calculation from Ref.~\cite{lyu2020high}, and consists with the result from exclusive electron scattering in Refs.~\cite{subedi2008probing,hen2014momentum}.

In previous $ (e,e^{\prime}) $ studies\cite{frankfurt1993evidence,egiyan2006measurement,fomin2012new}, at the region $ 1.5<x_{B}<2 $ the strengths of \textit{NN}-SRC are also extracted to be $ 5.2 $ for $ ^{56} $Fe and $ 5.3 $ for $ ^{27} $Al from the plateau of the ratios of $ (e,e^{\prime}) $ cross sections between heavy nuclei and the deuteron. The method of extracting the strengths of \textit{NN}-SRC from the plateau can remove interference of the mean-field part. However, this region $ 1.5<x_{B}<2 $ corresponds to a narrow range of low energy transfer, as shown in the shaded area of Fig.~\ref{fig4}(b) and  Fig.~\ref{fig4}(c). And at kinematics  $ Q^{2}>2{\rm\, GeV^{2}} $, the QE peak and $ \Delta $ production peak are not obvious, and the deep inelastic scattering (DIS) plays a dominate role in total $ (e,e^{\prime}) $ cross sections, which is sensitive to the internal structure of a proton or neutron. In  Fig.~\ref{fig4} of this paper, the $ Q^{2} $ region $ 0.5{\rm\, GeV^{2}} \sim 1 {\rm\, GeV^{2}}$ is selected for different nuclei. One can see that at this kinematics, the inclusive cross sections are mainly from the contributions of QE process and $ \Delta $ production process, which can better reflect the information of \textit{NN}-SRC effects and mean-field structure in nuclei.

~\

\emph{Summary. } The short-range correlation is an important nuclear property associated with the high-momentum components of the nuclear momentum distributions. In this paper, the effects of \textit{NN}-SRC on the inclusive electron scattering process are systematically studied, where the nuclear spectral functions are obtained from the combinations of the deformed RMF model and the LFD method. By using the appropriate form factors, and the decay width, we improve the theoretical cross sections in the $ {\Delta} $ production region.

We further separate $ (e,e^{\prime}) $ cross sections into MF and the \textit{NN}-SRC parts. One can see that the increase of \textit{NN}-SRC strength causes a reduction of the cross sections near the QE and the $ \Delta $ production peaks. By the method proposed in this paper, the strength of \textit{NN}-SRC for the $ ^{16} $O, $ ^{27} $Al, $ ^{56} $Fe, and $ ^{186} $W are constrained from the experimental data at sutible  kinematics to be $C _{A}=4.2$, $ 5.3$, $ 4.5 $, and $ 5.0 $, respectively, which correspond the proportions of high-momentum nucleons are $ Y=22.4\%$, $ 25.5\%$, $ 21.4\% $, and $20.2\%$.

The research in this paper can help to distinguish the nuclear effects and beyond-the-standard-model effects in neutrino-nucleus scattering experiments. Furthermore, the studies are useful for understanding the nuclear structure, and helpful for comprehending the strong interactions.

~\

\emph{Acknowledgements. }
This work was supported by the National Natural Science Foundation of China (Grants No. 11505292, No. 11822503, No. 11975167, No. 11961141003, and No. 12035011), by the Shandong Provincial Natural Science Foundation, China (Grant No. ZR2020MA096), by the Open Project of Guangxi Key Laboratory of Nuclear Physics and Nuclear Technology (Grant No. NLK2021-03), and by the Key Laboratory of High Precision Nuclear Spectroscopy, Institute of Modern Physics, Chinese Academy of Sciences(Grant No. IMPKFKT2021001).

\bibliographystyle{unsrt}
\bibliography{reference}

\end{document}